\documentclass[a4paper,11pt]{article}
\usepackage{jinstpub}
\pdfoutput=1 % if your are submitting a pdflatex (i.e. if you have
\title{\boldmath Exploring the lifetime and cosmic frontier with the MATHUSLA detector}

\author[a]{Cristiano Alpigiani}
\affiliation[a]{University of Washington, Seattle}

\emailAdd{Cristiano.Alpigiani@cern.ch}

\abstract{
The MATHUSLA detector to be installed on the surface above and somewhat displaced from the CMS interaction point (IP) will cover an area of $100\times 100$ m$^2$ containing many layers of scintillators planes to establish the space and time coordinates of charged particle tracks. This is an unprecedented detector in terms of size and continuous sensitivity over an area of $10^4$ m$^2$. This document describes the present MATHUSLA detector concept that is sensitive to both long-lived particles produced in the LHC collisions in CMS and cosmic ray extended air showers (EAS).  The ability to improve significantly cosmic ray studies by adding a $10^4$ m$^2$ layer of RPCs that have both digital and analogue readout similar to the ARGO-YBJ experiment will be discussed with focus on large zenith angle EAS. 
}

\keywords{Large detector systems for particle and astroparticle physics; Particle tracking detectors}

\collaboration[a]{on behalf of MATHUSLA collaboration}

\proceeding{XV Workshop on Resistive Plate Chambers and Related Detectors RPC2020\\
  10-14 February 2020\\
  Rome, Italy}

\begin{document}

\maketitle

\section{Introduction}

Long-lived particles (LLPs) occur in many extensions to the Standard Model (SM) with lifetimes that can be as long as the Big Bang Nucleosynthesis (BBN) bound of about $c \tau \lesssim 10^7$--$10^8$~m~\cite{BBNlimit}.  
Examples of models where such particles are predicted which can be produced at the Large Hadron Collider (LHC) include: Supersymmetric (SUSY) models such as RPV SUSY~\cite{Barbier:2004ez} and Stealth SUSY~\cite{Fan:2011yu}, models addressing the hierarchy problem such as Hidden Valleys~\cite{StrasslerA}, and models addressing dark matter \cite{DM1}. 

The main experiments at the LHC have extensive programs to search for such particles, covering lifetimes from a few centimeters to tens of meters. Searches for LLPs decaying into final states containing jets were carried out at the Tevatron ($\sqrt{s}=1.96$~TeV) by both the CDF~\cite{CDF} and D0~\cite{D0} Collaborations, at the LHC by the ATLAS and LHCb Collaborations in proton--proton collisions at $\sqrt{s}=7$~TeV~\cite{HVAtlas,LHCb}, by the ATLAS, CMS and LHCb Collaborations at $\sqrt{s}=8$~TeV~\cite{HVAtlas8TeV,SUSY-2014-02,CMSLLP,LHCb-PAPER-2016-065,LHCb-1612.00945} and more recently by the CMS and ATLAS Collaborations at $\sqrt{s}=13$~TeV~\cite{CMS-EXO-16-003,ATLAS13TeV,CalRatio}. 
To date, no search has observed evidence of beyond the Standard Model, neutral LLPs. However, their reach is limited by different factors such as the trigger, the presence of backgrounds from the collision or beam effects, and ultimately by the size of the detectors. For example, searches for LLPs decaying to hadrons (leptons) with less than a few 100~GeV ($\sim$~10~GeV) of visible energy in the event have particularly low trigger efficiency and are highly constrained by QCD and other backgrounds. These limitations could risk missing a discovery should LLP with a lifetime close to the BBN be created at the LHC collisions.

MATHUSLA (MAssive Timing Hodoscope for Ultra-Stable neutraL pArticles)~\cite{Chou:2016lxi} is a proposed large-scale surface detector to be located above ATLAS~\cite{ATLAS} or CMS~\cite{CMS} to study LLP produced by the High-Luminosity LHC (HL-LHC)~\cite{HL-LHC}. The $\sim$~90 m of rock between the interaction point (IP) and the detector's decay volume on the surface gives enough shielding for MATHUSLA to work in a clean environment. Being a background-free experiment increases the sensitivity to LLP lifetimes up to lifetimes of $10^{7}$ m and extends the sensitivity of the main detectors by orders of magnitude. LLP decays would be reconstructed as displaced vertices of upwards traveling charged particles. As a secondary physics objective, MATHUSLA would also be able to perform cosmic-ray physics measurements and help solve important puzzles in astroparticle physics.

A white paper describing the need for a detector like MATHUSLA was published by a large number of experimentalists and theorists in 2018~\cite{Curtin:2018mvb}. The MATHUSLA collaboration has made significant progress on the detailed background and design studies and presented a Letter of Intent~\cite{Alpigiani:2018fgd} to the LHCC. A test stand with a detector layout similar to the one envisioned for the MATHUSLA detector was assembled in the surface above ATLAS and took data during 2018. It was composed of one external layer of scintillators in the upper part and one in the lower part with six layers of RPCs (from the ARGO-YBJ experiment) between them. The overall structure was $\sim 6.5$ m tall, with a base of $\sim 2.9 \times 2.9$ m$^2$. The analysis of the test stand data has just been published~\cite{TSPaper}. The results provided empirical information on backgrounds coming from the LHC as well as from cosmic rays (CR). The collaboration is now seeking to construct a MATHUSLA demonstrator detector unit by 2021. The full-scale detector could become operational by 2025--26.

%%%%%%%%%%%%%%%%%
\section{The MATHUSLA detector}
%%%%%%%%%%%%%%%%%

%%%%%
\subsection{Basic Detector Principles}

\begin{figure}[hbtp!]
\begin{center}
\includegraphics[width=0.48\textwidth]{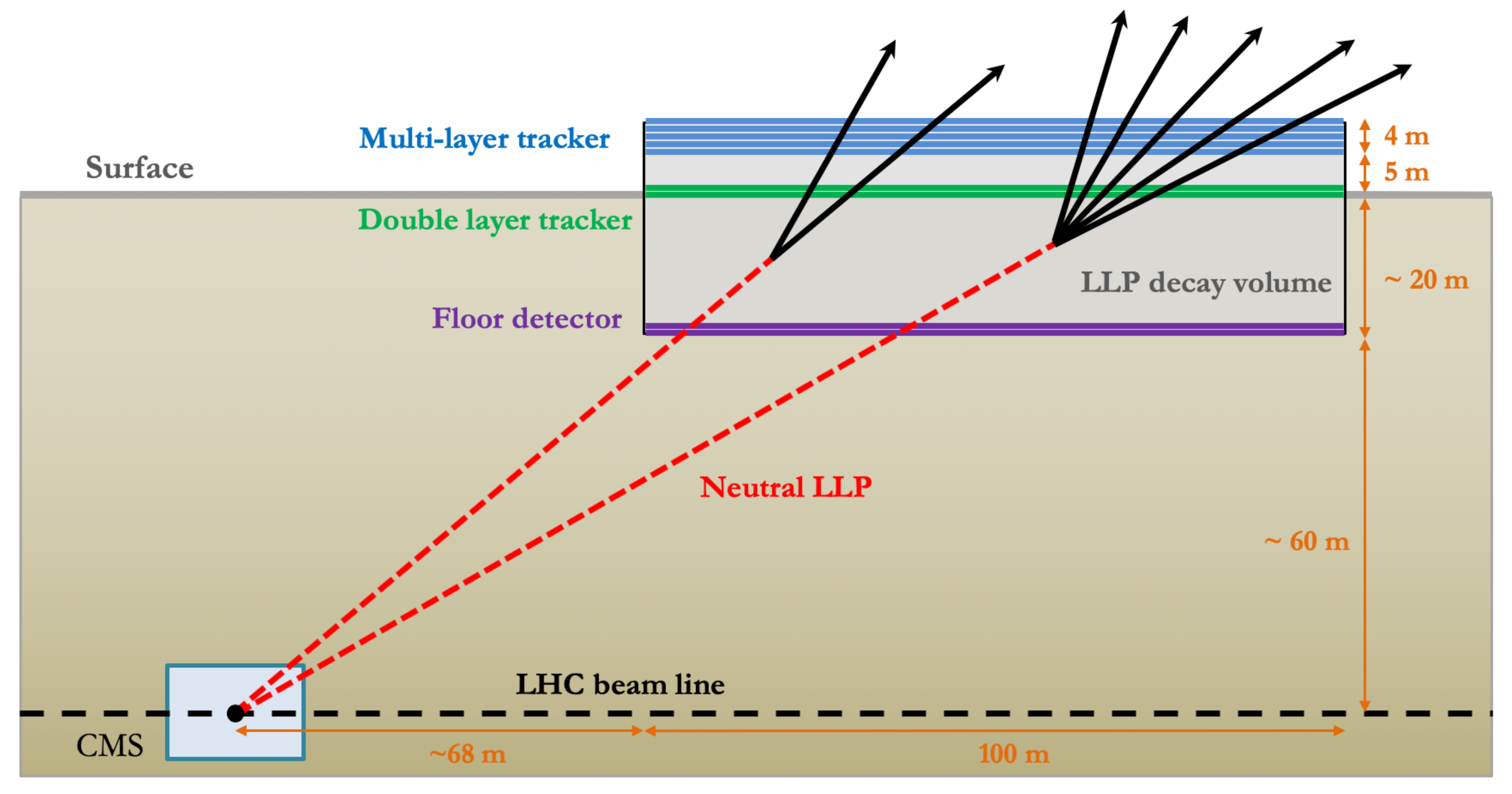}
\includegraphics[width=0.44\textwidth]{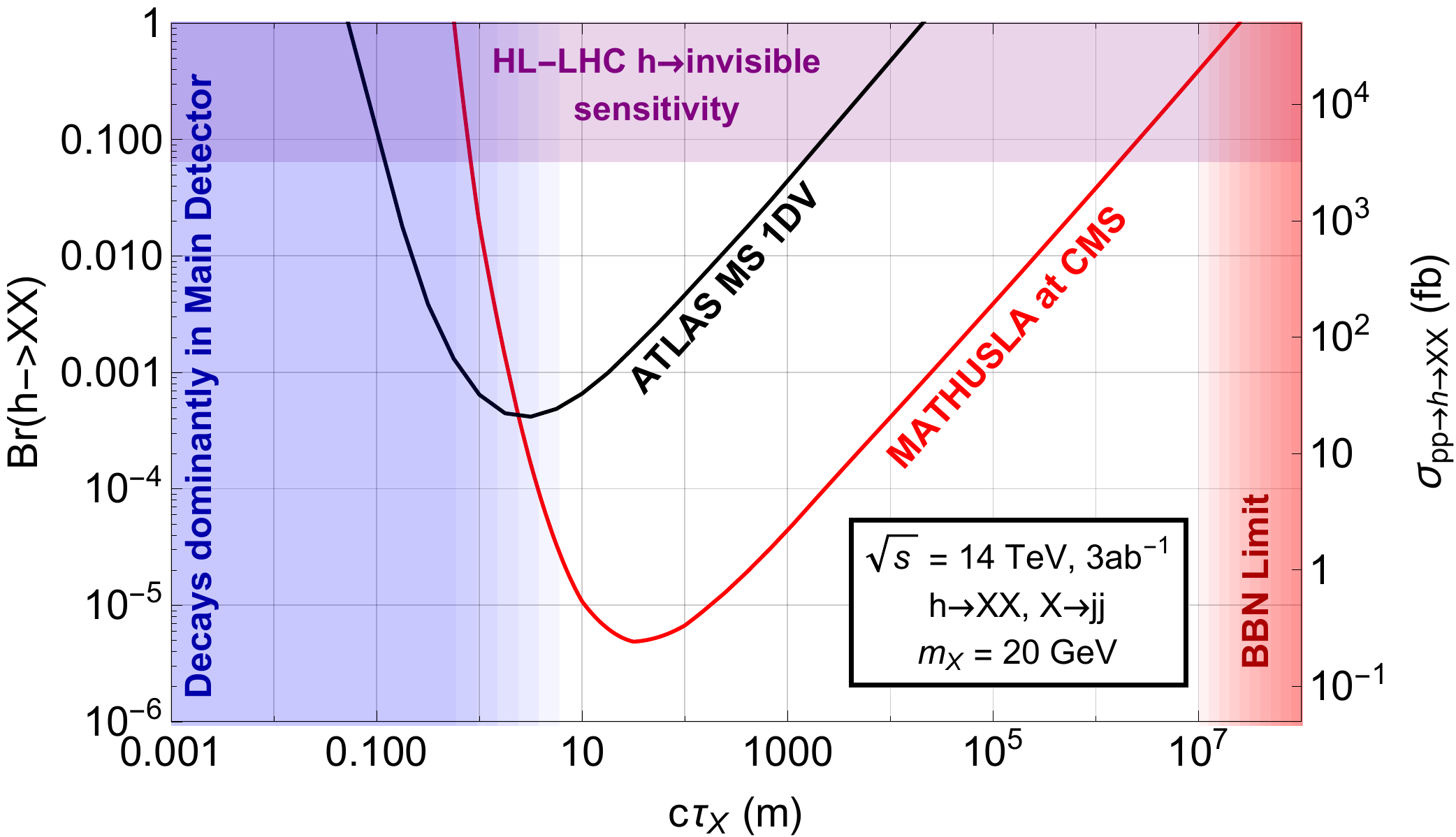}
\vspace{-0.5 cm}
\end{center}
\caption{Left: simplified MATHUSLA layout. Right: sensitivity comparison for a SM-like Higgs boson producing two long-lived scalars with a mass of 20 GeV decaying into hadronic jets, for MATHUSLA  in red and the ATLAS exclusion projection using a search in the Muon System~\cite{1MSVx}. The MATHUSLA curve corresponds to 4 LLPs decaying in the detector volume in a zero-background regime for a total integrated luminosity of 3 ab$^{-1}$ expected over the entire HL-LHC data taking period.
}
\label{fig:geometry}
\end{figure}

LLP near the BBN lifetime bound arising from exotic Higgs decays could be discovered if the detector had a linear size of $\sim\,$20~m in the direction of travel with good geometric coverage ($\sim$~5\% of solid angle), which results in a detector with linear dimensions of $\mathcal{O}$(100~m). The $\sim 90$ meters of rock between the collision point and the surface eliminates most backgrounds associated with $pp$ collisions.  Still, a large background of cosmic muons and backgrounds from high energy muons and neutrinos coming from the IP must be rejected which requires good tracking and vertexing capabilities. The proposed detector, illustrated in the left panel of figure~\ref{fig:geometry} along with two possible displaced vertices from LLP decays, is a large box of $100 \times 100 \times 25~{\rm m^{3}}$ volume, with a robust tracking system on its upper part, 25 m air decay volume and a tracking veto on the floor. An additional double-tracking layer 5 m below the main tracking system allows to enhance the particle position measurement precision close to the floor.

The expected sensitivity for a SM-like Higgs boson producing two long-lived scalars decaying into hadronic jets for the current benchmark geometry in the expected luminosity for the HL-LHC is shown in the right plot of figure~\ref{fig:geometry}. The MATHUSLA limit is obtained assuming 4 LLPs decaying in the detector volume in a zero-background regime. It is compared to the exclusion projection for a single displaced vertex search in the ATLAS Muon System considering the background expected at the HL-LHC~\cite{1MSVx}.

The dominant background comes from CRs, with a rate in the MHz range. Their rejection depends on the robust ceiling tracking system, with spatial and temporal resolutions in cm and nanosecond range, respectively. If the tracking layers span a vertical distance of a few meters, full 4-dimensional track and displaced vertex reconstruction is possible, which significantly reduces the combinatorial backgrounds as tracks must intersect in both space and time to form a vertex. Both Resistive Plate Chambers (RPCs) and plastic scintillators are time-tested technologies that meet the needed specifications. Since CRs travel downwards and do not inherently form displaced vertices, this signal requirement is expected to allow MATHUSLA to reach the near-zero-background regime.
 
The expected rate of muons from the HL-LHC collisions is $\mathcal{O}$(1~Hz). They are upwards traveling muons that do not generally produce a displaced vertex and that can be vetoed by the floor tracker. Upward going atmospheric neutrinos are estimated to be of order 10 to 100 per year, most can be rejected using the time of flight information of the LLP decay products. Moreover, they can be measured when there are no LHC beams. Neutrinos from LHC collisions are a subdominant background, estimated to be a few events during the entire HL-LHC data taking, and can be rejected with geometrical cuts and timing vetoes on non-relativistic charged tracks associated with the scattering event. A more detailed description of the possible backgrounds can be found in ref.~\cite{Alpigiani:2018fgd}, while more precise rate estimates from the analysis of the data collected by the test stand are described in ref.~\cite{TSPaper}.

\begin{figure}[hbtp]
\begin{center}
\includegraphics[width=0.44\textwidth]{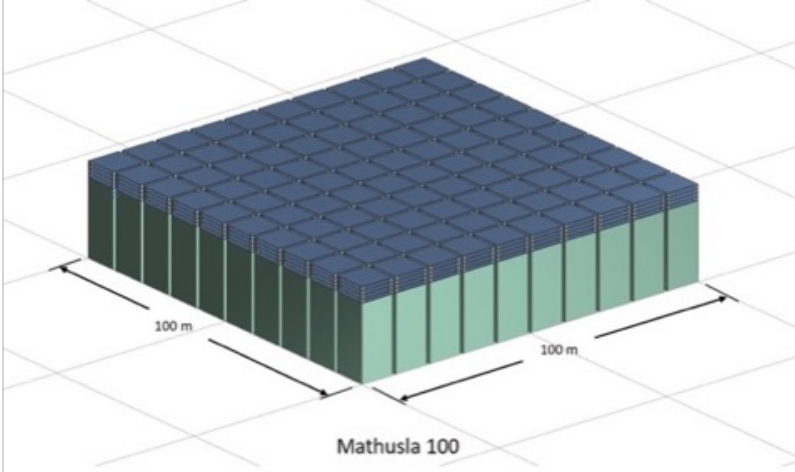}
\includegraphics[width=0.50\textwidth]{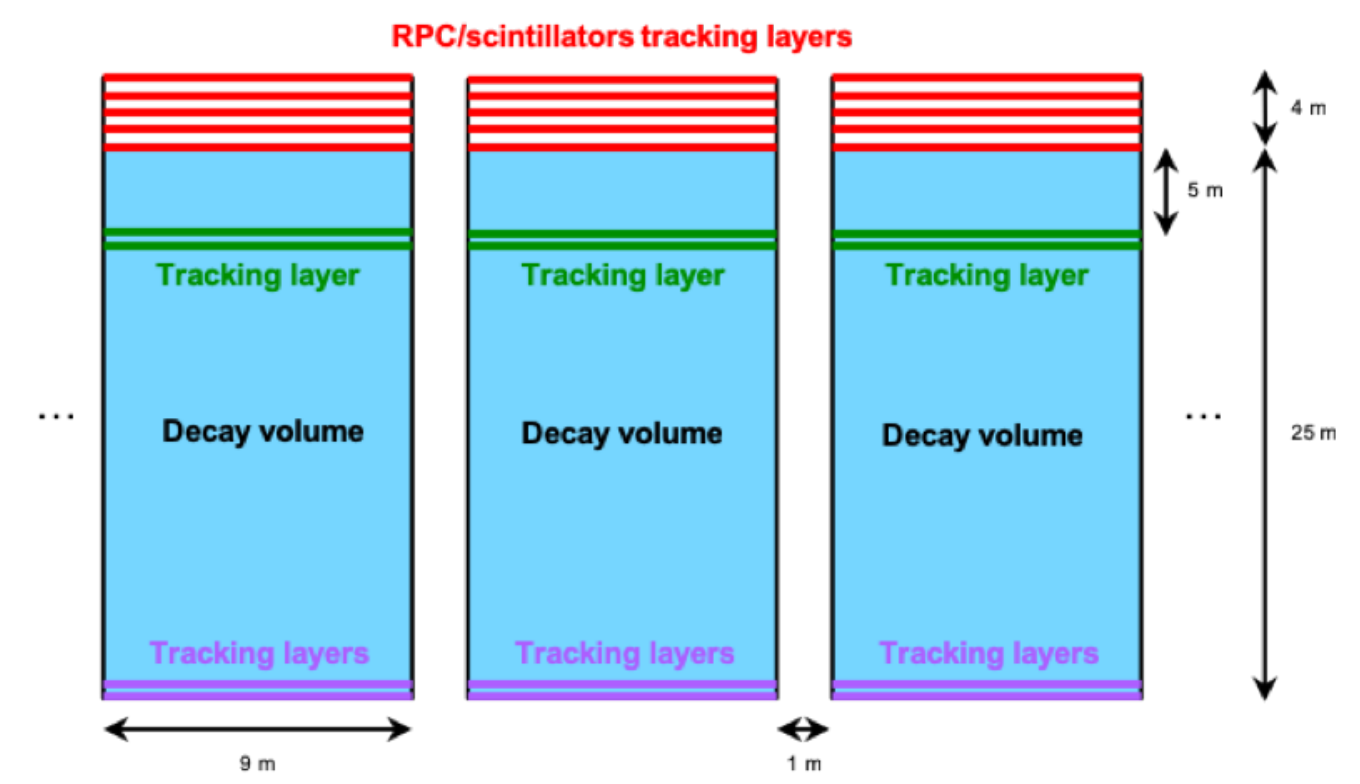}
\vspace{-0.5 cm}
\end{center}
\caption{
Schematics of the modular design of MATHUSLA (left) and structure of the modules (right).
}
\label{f.modular}
\end{figure}

%%%%%
\subsection{The modular concept}

MATHUSLA is designed to be a large area detector, requiring to cover a wide surface with detector material. Building MATHUSLA as an array of independent modules makes it a flexible and scalable detector, easy to adapt to the available land and the specific site conditions. It also allows for a staged integration with an incremental ramp-up. One of the advantages of MATHUSLA is that it is entirely parasitic: its construction and operation are not expected to interfere with the operation of ATLAS or CMS and its staged construction can happen as a completely independent plan from that of the HL-LHC and the experiments upgrade work.

The current design considers individual modules with a volume of $9\,\,\mathrm{m} \times 9\,\,\mathrm{m} \times 25\,\,\mathrm{m}$ and a separation of $\sim$ 1 m between modules as shown in figure~\ref{f.modular}. Each module includes two tracking layers on the floor to act as a veto for charged particles from the IP, an air-filled decay volume of 25 m and 5 tracking layers on the ceiling for track and vertex reconstruction. The extension of the decay volume by a few meters above ground has been decided as well as the inclusion of two extra tracking layers near the upper part of the decay volume to improve tracking and vertexing resolutions.
Studies conclude that a scintillator veto surrounding the entire volume is not needed. 

The baseline trigger system is driven by units of $3 \times 3$ modules, a choice based on the largest possible inclination angle for MATHUSLA which would be a very safe option for a $100~\rm {m} \times 100~\rm {m}$ detector.
The strategy is to collect all detector hits with no trigger selection and separately record the trigger information. The data rate is dominated by CRs, 1/(cm$^2$ minute), which gives $\sim 2$ MHz total rate for the main detector. Considering a unit of 9 m$^2$, two hits per module with 4 bytes per readout, the readout of 9 layers gives $\sim 1$ MB/s per unit. The readout will be based on ASIC chips with a cost target of 1 EUR per channel. The trigger is recorded separately and used for connecting to the CMS detector bunch crossing. By correlating MATHUSLA and CMS detector data, the LLP production mode and mass can be determined or at least constrained; the LLP boost distribution is tightly correlated with LLP mass once a production process is assumed, as demonstrated in ref.~\cite{Curtin:2017izq}. Conservatively assuming a distance from the IP of 150 m and an LLP with $\beta=0.7$, an optical fibre transmission to CMS with v$_\textrm{fibre} =0.5\mu s/100$ m will guarantee the detector around $3.25\mu$s or more to form trigger and get information to CMS Level-1 trigger (at the HL-LHC the CMS trigger will have a latency time of around 12 $\mu$s).

%%%%%
\subsection{Current geometry proposal}

\begin{figure}[hbtp]
\begin{center}
\includegraphics[width=0.36\textwidth]{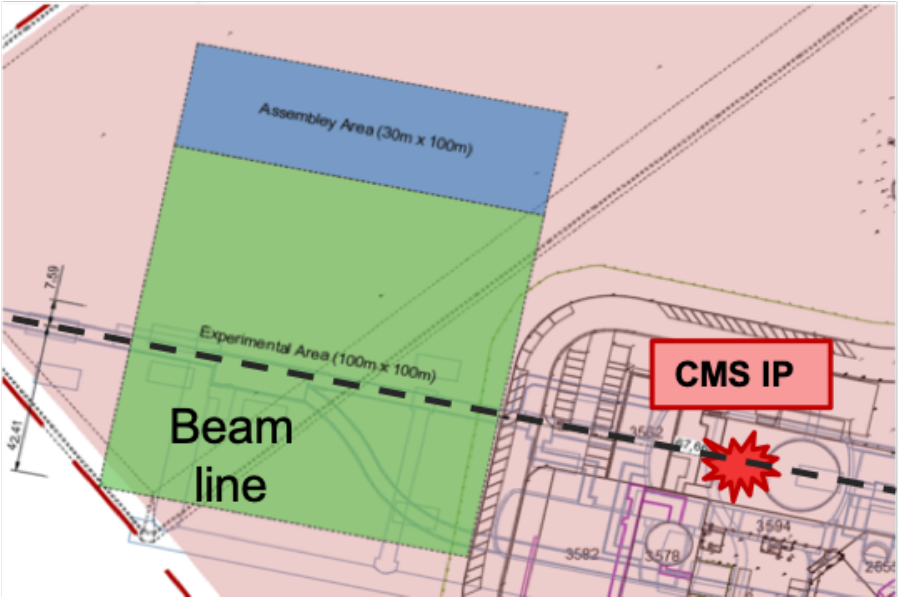}
\includegraphics[width=0.36\textwidth]{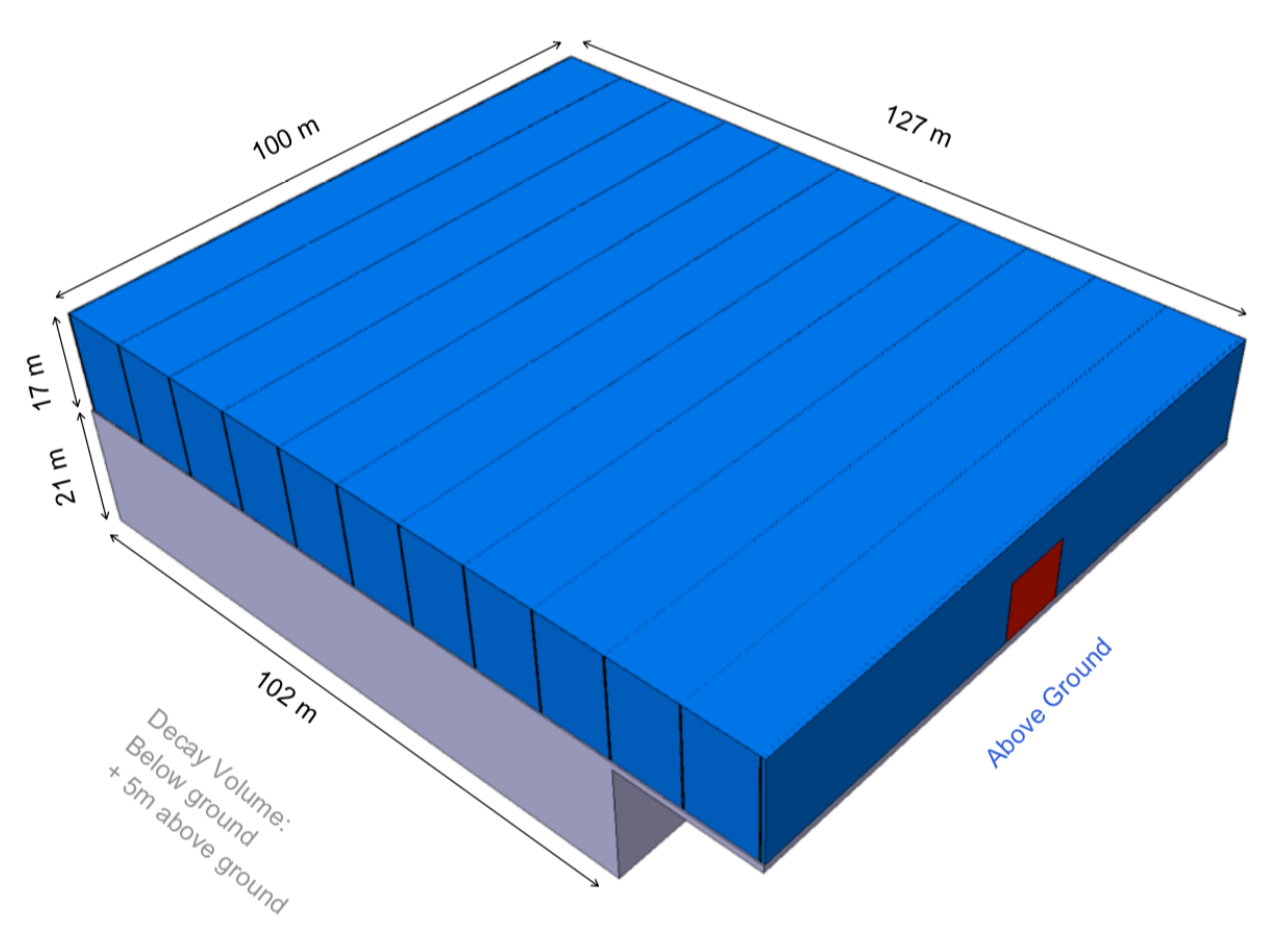}
\vspace{-0.5 cm}
\end{center}
\caption{
Left: CERN-owned land near CMS and location of MATHUSLA. The IP in CMS is marked as a red star. Right: Schematics showing the currently proposed dimensions of the MATHUSLA building.
}
\label{f.CMSsite}
\end{figure}

CERN owns an available piece of land near CMS that would be a suitable site for the detector~\cite{Alpigiani:2018fgd}. The MATHUSLA collaboration is working with Civil Engineers from CERN to define the building and the layout of the detector. Figure~\ref{f.CMSsite} shows the details on the planned position (left) and the size of the detector (right). 

The current proposal contemplates a 100 m $\times$ 102 m experimental area located on the surface of CMS together with a 30 m $\times$ 100 m adjacent area for the detector assembly. The total height of $\sim 40$ m includes a $\sim 25$ m decay volume, 21 m of which would be excavated, and 12 m in the upper part to host the tracker and the cranes' system for assembly and maintenance. Having a large part of the decay volume underground brings it closer to the IP, which increases the solid angle in the acceptance for LLPs generated in the collisions. To adjust to the available land, this proposal has a 7.5 m offset to the centre of the beams. The site allows for the detector to be as close as 68 m away from the IP.

The original MATHUSLA proposal~\cite{Chou:2016lxi} assumed a distance of 100 m from the IP both horizontally and vertically. Reducing these distances as explained above, the current proposal can reach a similar LLP sensitivity as the original detector with a final detector design more optimised, with smaller geometry that is cost-efficient and tailored to the available experimental site.

%%%%%
\section{MATHUSLA as a cosmic-ray telescope} 

MATHUSLA has all the qualities needed to act as an excellent CR telescope. \mbox{MATHUSLA's} large area provides good efficiency for extended air showers from primary CRs. Its combination with high-resolution directional tracking and proximity to ATLAS or CMS for correlated shower core measurements could allow more detailed studies of the core structure, crucial to determine the atomic number of the primary cosmic particles. These measurements, which do not interfere with the primary goal of LLP discovery, represent a ``guaranteed physics return'' on the investment of the detector, as well as an opportunity to establish a CR physics program. The qualitative CR physics case was discussed in \cite{Curtin:2018mvb}.

CRs up to the knee (3--4 10$^{15}$ eV) originate in supernova remnants and are accelerated by the 1st order Fermi mechanism in shock waves. The evolution of light nuclei spectra (p+He) could be an indication of the contribution of different populations of CRs (coming from different chemical compositions). Around the knee, CR measurements are performed through EAS arrays experiments. It is still not clear whether the mass of the knee is due to $p$ and He spectra or higher nuclei, and the results of the current experiments show some disagreement in this region. Therefore, analyse the primary proton spectrum is crucial to understand CR acceleration and propagation in the Galaxy, and precise measurements of the flux could allow to calculate the rate of secondary CR and atmospheric neutrinos.

\begin{figure}[hbtp]
\begin{center}
\includegraphics[width=0.8\textwidth]{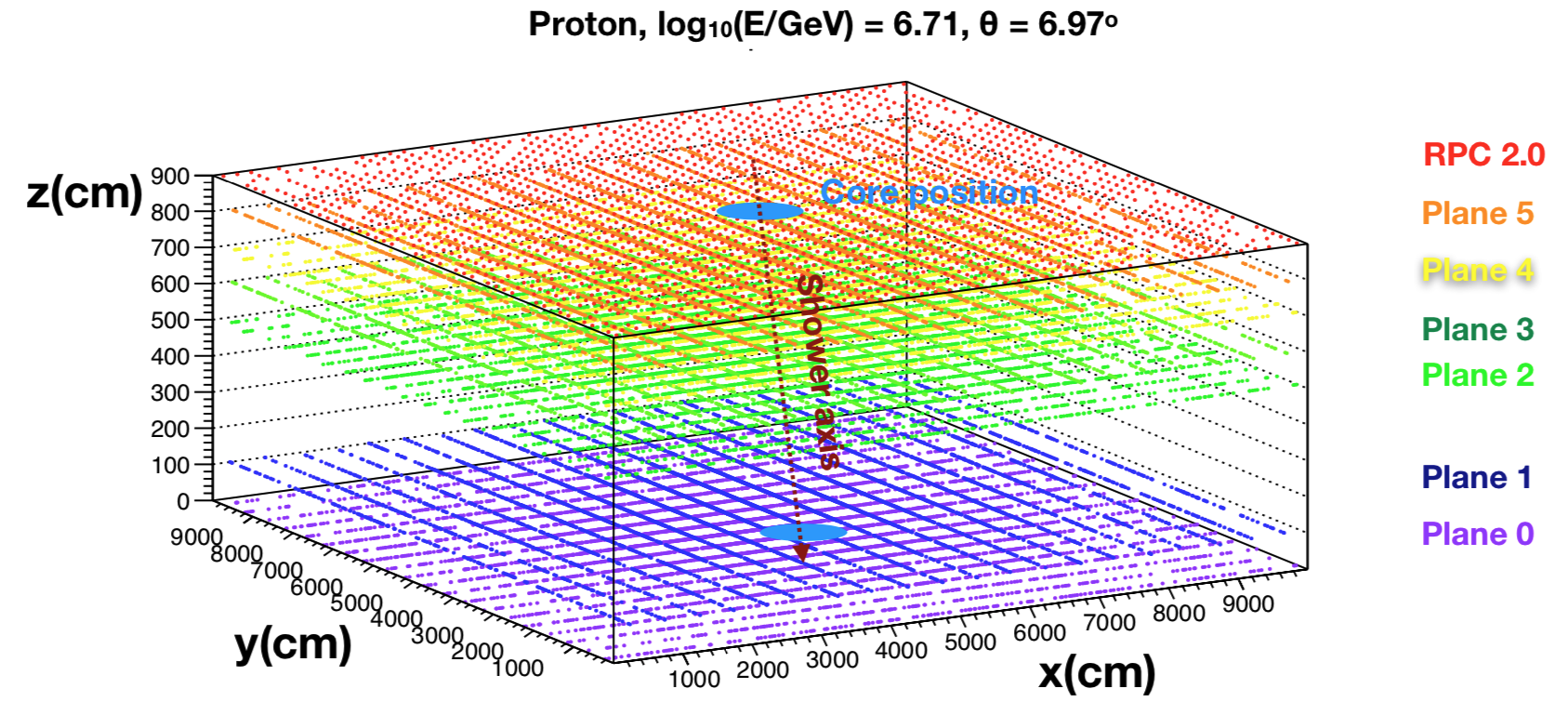}
\end{center}
\caption{
Example of a vertical air shower in MATHUSLA.
}
\label{f.CRShowerEvent}
\end{figure}

Figure \ref{f.CRShowerEvent} shows an example of a vertical shower reconstructed by the MATHUSLA detector using 7 layers of scintillator detectors (4 cm $\times$ 5 m bars) and an additional layer of RPC on top providing both analogue and digital measurements. The RPC can measure up to $10^4$ particles/m$^2$, and they are subdivided into Big Pads (BP). The position and arrival times of the particles can be measured with individual strips in each BP. For the current simulations, the BP provides space-time data of the shower with a time resolution of 1 ns and can have different sizes ($2\times5$ m$^2$, $2\times2$ m$^2$, $1\times1$ m$^2$). The detector is considered 100\% efficient and the signal is proportional to the particle density (linear response) assuming identical sensitivity to all charged particles.

The simulation is performed using CORSIKA 7.6400. For the high-energy ($E_h > 200$ GeV) and low-energy hadronic models, QGSJET-II-04 and Geisha are used, respectively. Only charged particles ($e^\pm$, $\mu^\pm$, $\pi^\pm$, $K^\pm$, $p^\pm$) are registered in each layer. Unstable particles are allowed to decay, but the decay products are removed from the simulation. The core of the air shower is scattered randomly in MATHUSLA, and for each shower event, the coordinates of the hit bars/BP and the arrival times of 1st particle to each hit bar/BP are registered.

The main disadvantages of using only scintillators are the impossibility of measuring the arrival times of the particles at the front of the EAS and the number of particles for events with zenith angles $\theta < 70^\circ$. This will not allow to measure the arrival direction of CRs with a high precision and it will make difficult to perform CR composition studies. 
%Moreover, we have no scale to calibrate primary energy of CRs since the umber of hits is saturated at HE's. 
The scintillator bar saturates at one hit making possible to find the impact point and arrival direction for one particle. For the current studies, we assign the coordinates of the centre of the bar to the hit, and register only the arrival time of the 1st particle that arrives to the scintillator bar.

The RPC digital readout might allow to improve the measurement of the spatial and temporal structure of an Extensive Air Shower (EAS), and perform low-density measurements. The analogue system has the advantage of allowing to measure the high density of particles up to $10^4$/m$^2$ in the streamer working mode, as shown by Argo-YBJ. In MATHUSLA, the RPCs will work in avalanche mode. This will extend of at least one order of magnitude the linearity range in the hit density measurement. This will allow to study the shower core profile with an unprecedented detail, thus expanding the measurements of CRs beyond the knee. The additional RPC layer could allow to precisely measure the shower front by having a good time-spatial determination of it. 

This could improve the determination of the core and the arrival direction of the shower, important for vertical EAS, where the saturation effects in the scintillation planes can lower the core and arrival direction precision. On the other hand, with measurements of the density of charged particles, the lateral distribution function (LDF) of charged particles can be obtained event-by-event, which can help to determine the energy scale of the primary CR and the composition of the CR nuclei. The energy scale can be estimated from the amplitude of the lateral distribution, and the primary composition could be studied by using the steepness of the LDF (the lighter and more energetic the air shower, the bigger the steepness of the LDF). Moreover, one RPC layer can improve the measurements of the vertical and inclined events on the energy and the deposited charge. All these additional information could allow the reconstruction of the all particle energy spectrum from vertical and inclined events up to 100 PeV, obtain large scale anisotropy maps in arrival directions of the CRs, measure small scale anisotropies in arrival directions, and search for point sources. Moreover, they will also allow testing, more precisely, the hadronic interaction models.

\begin{figure}[hbtp]
\begin{center}
\includegraphics[width=0.4\textwidth]{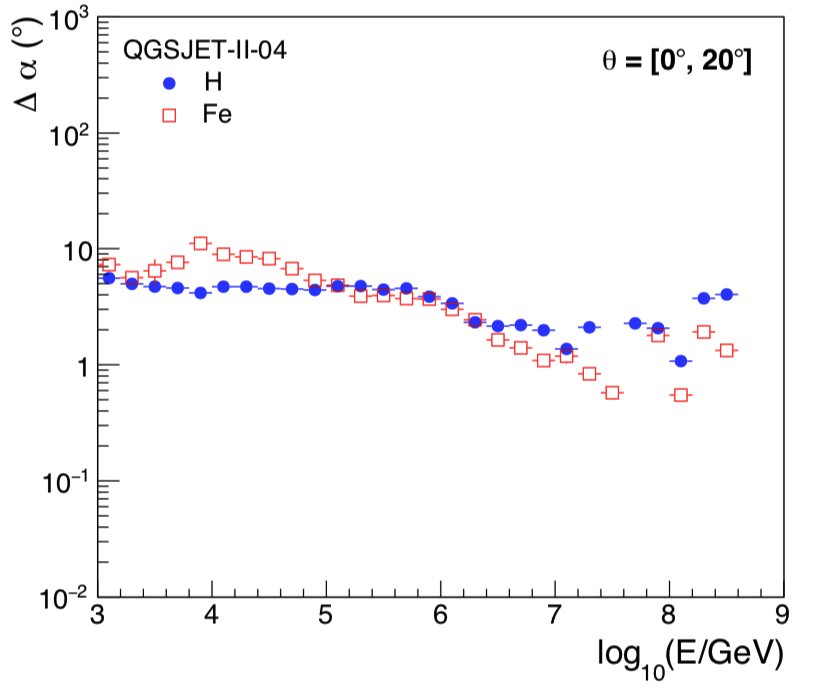}
\includegraphics[width=0.4\textwidth]{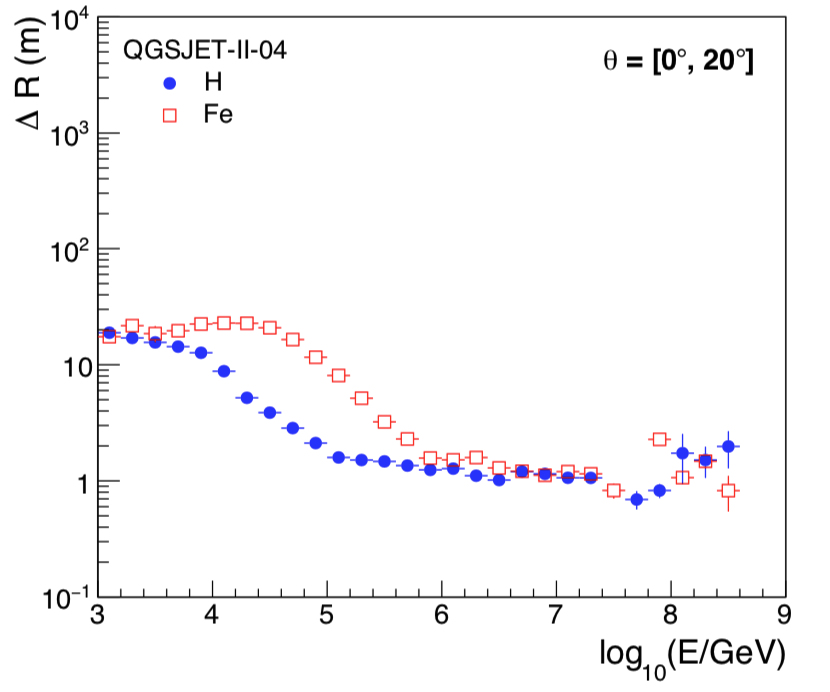}
\end{center}
\caption{
Preliminary performance (still to be optimised) for measurements of the arrival direction and core position for vertical events (number of hits > 50 and $\theta<20^\circ$) using information from scintillator and RPC detectors. 
}
\label{f.ArrivalCoreMeas}
\end{figure}

Figure \ref{f.ArrivalCoreMeas} shows the preliminary performance (no optimisation is performed yet) for measurements of the arrival direction and core position for vertical events (number of hits > 50 and $\theta<20^\circ$) using information from scintillator and RPC detectors. For comparison, the Argo-YBJ angular accuracy was lower than 6$^\circ$ in the TeV range.

%%%%%
\section{Conclusions}

LLPs occur in a wide variety of beyond the Standard Model scenarios addressing the most fundamental mysteries of particle physics. In this document, I presented the MATHUSLA detector that could extend the sensitivity to long decay lifetimes by orders of magnitude compared to LHC detector searches. Moreover, MATHUSLA could act as a CR telescope, and it could perform very precise CR measurements up to the PeV scale. By integrating a device with the possibility to measure arrival times and particle densities of extensive air showers, as an RPC, MATHUSLA can be employed as a CR detector and monitor a big portion of the sky above ($\theta < 80^\circ$), without limitation to inclined events.

\end{document}